\documentclass[prd,twocolumn,showpacs,floats,floatfix,nofootinbib]{revtex4}
\usepackage{graphicx}
\usepackage{dcolumn}
\usepackage{bm}
\def\spose#1{\hbox to 0pt{#1\hss}}
\def\lta{\mathrel{\spose{\lower 3pt\hbox{$\mathchar"218$}}
     \raise 2.0pt\hbox{$\mathchar"13C$}}}
\def\gta{\mathrel{\spose{\lower 3pt\hbox{$\mathchar"218$}}
     \raise 2.0pt\hbox{$\mathchar"13E$}}}
\newcommand{\be}{\begin{equation}}
\newcommand{\en}{\end{equation}}

\newcommand{\bea}{\begin{eqnarray}}
\newcommand{\ena}{\end{eqnarray}}

\newcommand{\cS}{c_{_{\rm S}}}
\newcommand{\cs}{c_{_{\rm S}}}

\newcommand{\epsm}{\epsilon _{{\rm m}}}

\newcommand{\Hu}{{\cal H}} 

\begin{document}

\title{Comment on ``Density perturbations in the ekpyrotic scenario''}

\author{J\'er\^ome Martin and Patrick Peter}
\email{jmartin@iap.fr, peter@iap.fr} 
\affiliation{Institut d'Astrophysique de Paris, GR$\varepsilon$CO,
FRE 2435-CNRS, 98bis boulevard Arago, 75014 Paris, France}
\author{Nelson Pinto-Neto}
\email{nelsonpn@cbpf.br} 
\affiliation{Centro Brasileiro de Pesquisas F\'\i sicas, 
   Rua Dr. Xavier Sigaud 150, Urca 22290-180, Rio de Janeiro, RJ, Brazil}
\author{Dominik J.~Schwarz}
\email{dschwarz@hep.itp.tuwien.ac.at}
\affiliation{Institut f\"ur Theoretische Physik, 
   Technische Universit\"at Wien,
   Wiedner Hauptstra\ss e 8--10, 1040 Wien, Austria} 

\date{\today}

\begin{abstract}
In the paper ``Density perturbations in the ekpyrotic scenario'',
it is argued that the expected spectrum of primordial perturbations 
should be scale invariant in this scenario. Here we show that, contrary 
to what is claimed in that paper, the expected spectrum depends on an 
arbitrary choice of matching variable. As no underlying (microphysical) 
principle exists at the present time that could lift the arbitrariness, 
we conclude that the ekpyrotic scenario is not yet a predictive model.
\end{abstract}
 \maketitle

\section{Introduction}

Recently, a model called the ekpyrotic scenario~\cite{ekp} was
proposed as a possible competitor to the inflationary paradigm~\cite{inflation}
on the basis that it could solve the problems of the hot big bang model 
and also produce an almost scale invariant spectrum of primordial scalar 
perturbations. Although the foundations of the model are still under 
debate~\cite{pyro}, we only shall comment on the 
controversial~\cite{Lyth,BF,HT} claim, crucial for the ekpyrotic 
scenario, that the power spectrum is scale-invariant~\cite{perturbekp}.

In Ref.~\cite{perturbekp}, the complicated five-dimensional evolution
of the background is modeled by an effective four-dimensional
spacetime which experiences a singular bounce at $\eta = 0$ ($\eta$
being the conformal time) as the effective scale factor vanishes. The
calculation of the power spectrum is a controversial issue as
different authors would use different matching
conditions~\cite{matching,MS,Ruth} at the singular bounce. There,
Bardeen's gravitational potential~\cite{Bardeen} $\Phi$ diverges. The
authors of Ref.~\cite{perturbekp} suggested to work with the
comoving density contrast $\epsilon_{\rm m}$ which is regular at $\eta
=0$ and has two linearly independent modes $D$ and $E$, such that
$\epsilon_{\rm m} = \epsilon_0 D(\eta) + \epsilon_2 E(\eta)$. At the
singular point the equality of the coefficients before and after the 
bounce is assumed \cite{perturbekp}: $\epsilon_0(0^+) = \epsilon_0(0^-)$ 
and $\epsilon_2(0^+) = \epsilon_2(0^-)$. Although this rule is given 
without justification (recall that the standard junction conditions 
follow from the Einstein equations), we shall refer to such equalities 
as matching conditions hereafter.

In Ref.~\cite{us} we criticized the matching conditions suggested in
Ref.~\cite{perturbekp}. We showed that they lead to an ambiguity, and
thus an arbitrary choice needs to be made that present-day physics
cannot make.  Ref.~\cite{perturbekp} was subsequently modified, see
Ref.~\cite{perturbekp2}, to address some critical comments that have
been made on the ekpyrotic model, one of them being the above
mentioned ambiguity exhibited in Ref.~\cite{us}. In the new
version~\cite{perturbekp2}, more details about the matching conditions
are given and one claim has been added (claim 1): ``{\it One situation of
special interest is ... where the potential is irrelevant at $\phi \to
- \infty$ and there is no radiation in the incoming state. In this
case, $\epsilon_2(0^-)=0$. In this case, we would obtain the same
final result from any matching rule which set $\epsilon
_2(0^+)=A\epsilon _2(0^-)$, with any constant $A$.}'' It concerns a 
particular case that we had
not considered and for which our result supposedly does not apply. It
should be noted however that the rest of Ref.~\cite{perturbekp2} 
inconsistently relies on the general case instead of this particular one, 
as it was the case in the preceding version~\cite{perturbekp}. 
Claim 1 clearly utilizes, in an essential manner, the ideas developed 
in Ref.~\cite{us} and represents a tentative {\sl a priori} response to 
the points Ref.~\cite{us} raised.

References \cite{perturbekp,perturbekp2} also state (claim 2):
``{\it The prescription is invariant under redefining the independent
solutions, e.g. by adding an arbitrary amount of the solution
$E(\eta)$ to $D(\eta)$. Matching any other non-singular perturbation
variable, defined to be an arbitrary linear combination of $\epsilon
_{\rm m}$ and $\epsilon _{\rm m}'$ with coefficients which are
non-singular background variables (defined to possess power series
expansions in $\tau $, as above) will, with the same prescription of
matching the amplitudes of both linearly independent solutions, also
yield precisely the same result}''. This is an important point in the
ekpyrotic scenario because, if true, it would support the use of the
variable $\epsilon _{\rm m}$ as a tool to define matching conditions.

Below we show that both quoted claims are incorrect.

\section{Is the Ekpyrotic spectrum unique?}

Let us start with claim 1, i.e. the question concerning the invariance
of the spectrum under rescaling of the parameter $\epsilon_2$ and let
us first define our notation. The authors of Ref.~\cite{perturbekp2}
consider the equation of state close to the singular bounce to be
$
\omega \equiv p/\rho = \omega_0 + \omega_1\eta + \omega_2\eta^2 
+ \cdots 
$.
Before the bounce, the kinetic energy of the scalar field dominates
and the dynamics can be approximated by that of a free scalar field
for which $\omega_0 = 1$ and $\omega_1 = \omega_2= 0$ (note that one
can show that the values of $\omega_i$, $\forall i>2$, are irrelevant
for this discussion). This should apply for $\eta <0$ close to the
singularity ($|\eta|\ll1$) and corresponds precisely to the situation
that is assumed in claim 1 and to the case where the equations of
Ref.~\cite{us} cannot be applied (as mentioned in that article), hence
the argument of Ref.~\cite{perturbekp2}. Below, we complete the
arguments of Ref.~\cite{us} to include this particular case and show
that the conclusions of Ref.~\cite{us} remain unchanged.

At leading order, the behavior of the scale factor is a simple
power-law,
$
a(\eta )=\ell_0(-\eta )^{1/2}/(2\sqrt{2})
$,
in agreement with the notation of Eq.~(24) in Ref.~\cite{us}. The
comoving Hubble rate and the sound speed are
$
{\cal H}\equiv a'/a = 1/(2\eta)$ and $\cS^2 = 1$.
The solution for the Bardeen potential~\cite{Bardeen,perturb} in the
long wavelength limit reads
\begin{equation}
\Phi = \frac{3}{4}B_1(k)\frac{{\cal H}}{a^2}
     + \frac{3}{4}B_2(k)\frac{{\cal H}}{a^2}
       \int^{\eta }\frac{{\rm d}\tau}{\theta^2},  \label{Phisol}
\end{equation}
with $\theta \equiv 1/(a\sqrt{1+\omega})$ and the integral sign stands
for the primitive of the integration kernel. The coefficients $B_1$
and $B_2$ are functions of the comoving wavenumber $k$, to be
determined by means of a matching with the initial vacuum
condition~\cite{perturbekp,us}. In the case at hand,
\begin{equation}
\Phi = - {3\over \ell_0^2\eta^2}B_1 + {3\over 8}B_2. \label{cool}
\end{equation}
Note that the equation of motion for the Bardeen potential for a free 
scalar field reads
$
\Phi'' + (3/\eta )\Phi' + k^2 \Phi = 0
$,
which can be solved exactly in terms of Bessel functions,
\begin{equation} 
\Phi = {3 \over (-k\eta)} 
\left[\frac14 B_2 J_1(-k\eta) + 
\frac\pi 2 {k^2\over \ell_0^2} B_1 N_1(-k\eta)\right],
\label{exact}
\end{equation} 
whose expansion in powers of $\eta$ permits to recover the special
solution~(43-44) of Ref.~\cite{perturbekp} as well as
Eq.~(\ref{cool}).

Before the bounce the density contrast on comoving hypersurfaces,  
$\epsilon_{\rm m} = - 2 k^2 \Phi/(3{\cal H}^2)$, is given by
\begin{equation}
\epsilon_{\rm m} =
{8 k^2 \over \ell_0^2} B_1 - k^2 \eta^2 B_2.
\label{epsexact}
\end{equation}
Comparison of Eq.~(\ref{epsexact}) with Eqs.~(43-44) of
Ref.~\cite{perturbekp} yields
\begin{equation}
\label{epbefore}
\epsilon^< _0(k) = {8 k^2 \over \ell_0^2} B^<_1, \quad 
\epsilon^< _2(k) = - k^2 B^<_2,
\label{Em}
\end{equation}
showing that the two ``modes'' decouple.
It is interesting to compare with the case $\omega_1^<\neq 0$, for
which one has~\cite{us} (here we assume $\omega_2^>=0$ for the sake of
simplicity; this does not change in any way the conclusion)
\begin{eqnarray}
\epsilon^<_0(k) &=& \frac{8k^2}{\ell_0^2} B^<_1
- \frac{128 \ln 2\, k^2}{9\omega_1^{<2}} B^<_2 , 
\label{E0m} \\ 
\epsilon^<_2(k) &=& \frac{9 k^2\omega_1^{<2}}{8\ell_0^2} B^<_1
- \left(1 + 2\ln 2\right)k^2 B^<_2.
\label{E2m}
\end{eqnarray}
Let us note at this point that, with the choice we made of the
normalization of the integral in Eq.~(\ref{Phisol}), the limit
$\omega_1\to 0$ is singular, as can be seen from
Eqs.~(\ref{E0m})~--~(\ref{E2m}) and as already mentioned in
Ref.~\cite{us}. Another choice is possible as one is free to add an
arbitrary constant amount of $B_2$ into $B_1$, simply by making the
integral a definite one.  In particular, one can arrange that the
constant factor $(\ln 2)$ in the above equations be made to vanish
from the outset, and it is easy to convince oneself that such a choice
does not modify anything if the correct equations are used.

In the ekpyrotic scenario, and in the long wavelength limit, the
coefficient $B_1^<\propto k^{-3/2}$ is dominant with respect to
$B_2^<\propto k^{-1/2}$. This means that the relation $\epsilon_2
\simeq 9 \omega_1^2 \epsilon_0/64$ holds true if $\omega_1\neq 0$,
whereas it does not if $\omega_1 =0$. In the latter case, one has
$\epsilon _2=-\ell _0^2B_2\epsilon _0 /(8B_1)\propto k\epsilon_0 \neq
0$, see Eq.~(\ref{Em}).
The authors of Ref.~\cite{perturbekp2} missed that the limit $\omega
_1 \rightarrow 0$ does in no way imply $\epsilon_2=0$ exactly, as
implied by claim 1, in which this value is said to be used to perform
the matching. Moreover, as demonstrated below, the fact that $\epsilon
_2$ becomes subdominant does not remove the ambiguity noticed in
Ref.~\cite{us}.

This is also related to the excessive claim in \cite{perturbekp2} that
``{\it there is no long wavelength contribution to $\zeta$ in the
collapsing phase}''. Inserting (\ref{exact}) into the gauge-invariant
variable
$
\zeta \equiv (2/3) (\Hu^{-1}\Phi' + \Phi)/(1
+ \omega)+ \Phi
$ (see e.g. \cite{MS,Bardeen,perturb}), gives
\begin{equation}
\zeta =\frac{1}{2}B_2J_0(-k\eta) + 
\frac{\pi}{4}\frac{k^2}{\ell_0^2}B_1N_0(-k\eta) \end{equation}
which, in the limit $\eta \to 0$, yields
\begin{equation}\zeta\sim \frac{1}{2}B_2 - 
\frac{1}{2}\frac{k^2}{\ell_0^2}B_1 [\ln(-k\eta) + \gamma_{_{\rm E}}],
\end{equation}
where $\gamma_{_{\rm E}}$ is Euler constant coming from the expansion
of the Bessel function. Since the second term is singular
the first term is subdominant, but that does not imply that it
vanishes altogether.

After the singular bounce ($\eta>0$), $\omega_1^> \neq 0$ and one
recovers the relations~\cite{us} 
\begin{eqnarray}
\epsilon^>_0(k) &=& - \frac{8k^2}{\ell_0^2} B^>_1
- \frac{128 \ln 2\, k^2}{9\omega_1^{>2}} B^>_2 , 
\label{E0p} \\ 
\epsilon^>_2(k) &=& - \frac{9 k^2\omega_1^{>2}}{8\ell_0^2} B^>_1 -
\left(1 + 2\ln 2\right)k^2 B^>_2.
\label{E2p}
\end{eqnarray}
The authors of Ref.~\cite{perturbekp2} propose to glue the epoch after
the bounce to the epoch before the bounce by imposing that
$\epsilon_0$ and $\epsilon_2$ are the same before and after the
singularity. Contrary to standard junction conditions, and until some
more microphysics is specified, this recipe does not rest on any
physical principle and one may wonder what is the rationale behind
it. The reason why one is pushed to adopt a new rule in this case is
linked to the fact that one is dealing with a perturbative approach
around a singular background whose meaning is
questionable~\cite{Lyth}. Moreover, we argued in Ref.~\cite{us} that
this prescription is ambiguous since an arbitrary rescaling factor
$f(\omega,k)$ shows up in the final result. Therefore, the scenario
contains an arbitrary function in a physically measurable quantity:
until this function can be calculated unambiguously by first
principles, the model cannot be falsified.

As in Ref.~\cite{us}, we rescale $\epsilon_2$ by the completely
arbitrary factor $1/f$.  This gives the constant part of the
gravitational potential
\begin{equation}
B_2^> = B_2^< {f^>\over f^<} + {9 \omega_1^{>2}\over 8 \ell_0^2} B_1^<.
\label{B2p}
\end{equation} 
This is the main equation of Sec.~II. It represents the equivalent of
Eq.~(54) of Ref.~\cite{us} for the case $\omega _1^<=0$. On the other
hand, the decaying mode amplitude reads
\begin{equation}
B_1^> = - (1+2\ln 2) B_1^< -{16 \ln 2\, \ell_0^2 \over 9 \omega_1^{>2}}
B_2^< {f^>\over f^<},
\end{equation} 
given here for the sake of completeness.

Note that if the expansion of the equation of state
parameter is done to higher order in $\eta$, all the relations of this
section are only changed by inclusion of $\omega_2$ through the
replacement
$ 
(3/8)\omega_1^2\to w^{(2)} \equiv \omega_2+(3/8)\omega_1^2
$, 
in agreement with
Ref.~\cite{perturbekp,perturbekp2}, while none of the $\omega_i$ for
$i>2$ does contribute.

Eq.~(\ref{B2p}) shows that the spectrum effectively acquires a scale
invariant piece $\propto B_1^<$, together with an arbitrary piece
$\propto B_2^< f^>/f^<$. Indeed, as discussed in Ref.~\cite{us}, the
functions $f^<$ and $f^>$ can depend on the background constants
(e.g., $\omega$) and on the considered scale $k$ in a way which is not
yet given from first principles. The choice $f^> = f^<$ fixes the
$k-$dependence of the spectrum, but, without physical justification,
this remains an arbitrary choice, equivalent to assume a scale
invariant spectrum from the outset.

\section{An example}

Let us now consider claim 2 according to which the choice of $\epsm$
is essentially unique. The ambiguity inherent to the method proposed
in Refs.~\cite{perturbekp,perturbekp2} can be most clearly exhibited by
means of an example suggested at the Euro-conference in Annecy in
December~2001 by G.~Veneziano~\cite{Ve1}. This section relies
completely upon his idea.

In Ref.~\cite{perturbekp2} it is argued that the density contrast
$\epsilon _{\rm m}$ is the quantity of interest because it is finite
at $\eta =0$ and we have shown above how this quantity is used to
propagate the spectrum through the singularity. However, $\epsilon
_{\rm m}$ is not the only finite, physically relevant, quantity. Based
on the method developed in Ref.~\cite{Ve}, we could equally well
consider and use the conjugate momentum $\Pi $ to the conserved
quantity $\zeta $~\cite{MS} which obeys the Hamilton like equations,
valid for isentropic perturbations,
\begin{equation}
\Pi=z^2\zeta ', \quad \Pi '=-k^2c_{_{\rm S}}^2z^2\zeta
,\label{PiHydro}
\end{equation}
where $z^2\equiv a^2(1+\omega )/c_{_{\rm S}}^2$. It is worth pointing
out that since the quantity $\zeta$ diverges at the singularity, it
may be more useful to combine Eqs.~(\ref{PiHydro}) into the single one
\begin{equation}
\Pi'' +\left[ 3\left( \cs^2 -\omega\right)
-2\right]\Hu \Pi' + k^2 \cs^2 \Pi =0.\label{PiSeul}
\end{equation}
In the neighborhood of the singularity, $\cs^2=\omega=1$, and the
solutions of Eq.~(\ref{PiSeul}) are $-k\eta J_1(-k\eta)$ and $-k\eta
N_1 (-k\eta)$, where $J_1$ and $N_1$ are Bessel function of the first
and second kind respectively. These solutions are
completely regular at $\eta=0$. Note that Eqs.~(\ref{PiHydro}) and
(\ref{PiSeul}) apply precisely in the two cases of interest here,
namely that of purely hydrodynamical perturbations and that of a free
scalar field, by replacing $\cS^2\to 1$~\cite{MS} in this last
case. The quantity $\Pi$ is related to $\epsilon _{\rm m}$ as
\begin{equation}
\label{bleue}
\Pi=a^2{\cal H}\epsilon _{\rm m}\, .
\end{equation}
Note also that the use of $\Pi$ may be argued to be more appropriate
in view of the fact that in the regular bounce case~\cite{bounce},
even though admittedly a different case as the ekpyrotic model, the
variable $\epsm$ ends up being an odd function of $\eta$, whereas
$\Pi$ is even. In the short-duration limit, that would imply that
$\epsm$ experiences a jump whereas $\Pi$ is continuous.

Since $|a^2\Hu| \sim \ell_0^2/16$ is finite at the singularity
$\eta=0$, this relation clearly shows that the variable $\Pi$, being a
linear combination of $\epsilon _{\rm m}$ and $\epsilon _{\rm m}'$
(with vanishing coefficient for the latter), satisfies all the
requirements of claim 2. Accordingly, it seems that there is no
convincing reason to use $\epsilon _{\rm m}$ rather than $\Pi$. In the
vicinity of the bounce, $\Pi $ reads
\begin{equation}
\label{decadix}
\Pi =\frac{\ell _0^2s}{16}\biggl[\epsilon _0-\frac{3}{4}\omega
_1\epsilon_0 \eta +\biggl(\epsilon _2 -\frac{3w^{(2)}}{8}\epsilon
_0\biggl)\eta ^2 +\cdots \biggr]\, ,
\end{equation}
where $s=-1$ before the bounce and $s=+1$ after the bounce. The
proposal of Ref.~\cite{perturbekp} then would consist in assuming that
the coefficients in front of the constant term on the one hand and in
front of the $\eta ^2$ term on the other hand are the same before and
after the bounce. In the present context, this reduces to
\begin{equation}
-\epsilon _0^< = \epsilon _0^> \, ,\quad 
-\epsilon _2^< = \epsilon _2^>-\frac{3w^{(2)>}}{8}
\epsilon _0^> \, ,
\end{equation}
since $w^{(2)<}=0$ by definition. From these relations, it is easy to
establish that
\begin{equation}
\epsilon _0^> = -\frac{8k^2}{\ell _0^2}B_1^<, \quad 
\epsilon _2^> = k^2B_2^< - \frac{3w^{(2)>}k^2}{\ell _0^2}B_1^< \, .
\end{equation}

Finally, the computation of $B_2^>$ shows that the spectrum no longer
contains a scale invariant piece,
\begin{equation}
B_2^>=-B_2^< \propto k^{-1/2}\, ,
\end{equation}
but gives a spectral index equal to $3$. It should be noticed that the
same calculation goes through, with the same resulting spectral index
if, in Eq.~(\ref{bleue}), one replaces the factor $a^2{\cal H}$ by its
magnitude $\vert a^2{\cal H}\vert $ which is not only regular but also
symmetric across the bounce. Therefore, the choice of the variable
used to apply the rule proposed in Ref.~\cite{perturbekp} plays a
crucial role. In the absence of an underlying reason to choose a
variable rather than another at the present stage, $\Pi$ seems to be
as relevant as $\epsm$, and we are led to the conclusion that claim 2
is definitely erroneous. Moreover, $\Pi$ is not the only well-defined
variable that leads to such a conclusion: there is an infinite number
of such possible combinations.

\section{Conclusions}

To conclude, let us mention yet another possibility. In
Ref.~\cite{Ruth}, it was suggested that the condition used in
Ref.~\cite{perturbekp} may be cast in the form of a matching of the
energy perturbation as seen by an observer comoving with the fluid,
and that such a matching is ``{\it at least as natural as matching the
energy in the longitudinal gauge}''. Firstly, one should notice that
no rigorous proof exists that the conditions of Ref.~\cite{Ruth} are
equivalent to the prescription utilized in
Ref.~\cite{perturbekp}. Secondly, we have shown in Ref.~\cite{us} that,
in the well known case of radiation to matter transition for which one
can compare with the exact solution, this matching condition yields an
incorrect result. A similar conclusion has been reached in
Ref.~\cite{BF} in the case of the reheating transition in which the
equation of state also jumps. However, it can be argued that a
bouncing situation is in no way comparable to a jump in the equation
of state and that the usual matching conditions may then not
apply. This is indeed what was found in Ref.~\cite{bounce} for the
case of a non singular bounce. The case for the ekpyrotic scenario is
still more involved, as it was also argued that the bounce must be
singular if the brane collision is to produce radiation~\cite{Syksy}.
In any case, matching through an unspecified bounce is still an open
question as there is no geometrical or physical argument imposing some
particular choice, and no conclusion can be made about the power
spectrum of perturbations in general, unless some non-singular bounce
is specified, as it is the case of Ref.~\cite{bounce}.

Finally, we would like to insist on the fact that the overall
gauge-invariant perturbation theory may be completely meaningless
when a singularity is reached, since some gauge transformations that
are admissible in any other situation could turn out to be
singular. In such a situation, a ``gauge-invariant'' variable will, at
the singular point, become ``gauge-dependent'', and its use rather
than the use of any other variable become an arbitrary, physically
meaningless, choice.  In fact, the transformation between the zero
shear (conformal Newtonian) slicing and the comoving slicing is indeed
singular at the bounce, which actually means that one or both slices
are unphysical. Lyth~\cite{Lyth} has shown that there is no slicing on
which the density contrast and the intrinsic curvature perturbation
are finite simultaneously. In such a situation, linear perturbation
theory becomes meaningless.
\vspace{0.15cm}
\par
\centerline{\bf Acknowledgments}
\par

We would like to thank Robert Brandenberger, Ruth Durrer, Justin
Khoury, David Lyth and Raymond Schutz for many enlightening
discussions. We are especially indebted with Gabriele Veneziano for
proposing the variable $\Pi$ of Sec.~III and numerous other
discussions. NPN would like to acknowledge IAP for warm hospitality
during the time this work was being done, and CNRS and CNPq for
financial support. DJS acknowledges financial support from the
Austrian Academy of Sciences.

\end{document}